\documentclass[aip,apl,reprint]{revtex4-1}
\usepackage[T1]{fontenc}
\usepackage[latin9]{inputenc}
\usepackage{amstext}
\usepackage{graphicx}
\usepackage{amssymb}
\usepackage{multirow}
\usepackage{dcolumn}
\usepackage{color}
\usepackage{amsmath}

\newcommand{\TIL}{{\raise.17ex\hbox{$\scriptstyle\mathtt{\sim}$}}}
\begin{document}

\title{Planar Superconducting Resonators with Internal Quality Factors above One Million}

\author{A. Megrant}
\affiliation{Department of Physics, University of California, Santa Barbara, California 93106-9530, USA}
\affiliation{Department of Materials, University of California, Santa Barbara, California 93106-9530, USA}
\author{C. Neill}
\affiliation{Department of Physics, University of California, Santa Barbara, California 93106-9530, USA}
\author{R. Barends}
\affiliation{Department of Physics, University of California, Santa Barbara, California 93106-9530, USA}
\author{B. Chiaro}
\affiliation{Department of Physics, University of California, Santa Barbara, California 93106-9530, USA}
\author{Yu Chen}
\affiliation{Department of Physics, University of California, Santa Barbara, California 93106-9530, USA}
\author{L. Feigl}
\affiliation{Department of Materials, University of California, Santa Barbara, California 93106-9530, USA}
\author{J. Kelly}
\affiliation{Department of Physics, University of California, Santa Barbara, California 93106-9530, USA}
\author{Erik Lucero}
\affiliation{Department of Physics, University of California, Santa Barbara, California 93106-9530, USA}
\author{Matteo Mariantoni}
\affiliation{Department of Physics, University of California, Santa Barbara, California 93106-9530, USA}
\affiliation{California NanoSystems Institute, University of California, Santa Barbara, California 93106-9530, USA}
\author{P. J. J. O'Malley}
\affiliation{Department of Physics, University of California, Santa Barbara, California 93106-9530, USA}
\author{D. Sank}
\affiliation{Department of Physics, University of California, Santa Barbara, California 93106-9530, USA}
\author{A. Vainsencher}
\affiliation{Department of Physics, University of California, Santa Barbara, California 93106-9530, USA}
\author{J. Wenner}
\affiliation{Department of Physics, University of California, Santa Barbara, California 93106-9530, USA}
\author{T. C. White}
\affiliation{Department of Physics, University of California, Santa Barbara, California 93106-9530, USA}
\author{Y. Yin}
\affiliation{Department of Physics, University of California, Santa Barbara, California 93106-9530, USA}
\author{J. Zhao}
\affiliation{Department of Physics, University of California, Santa Barbara, California 93106-9530, USA}
\author{C. J. Palmstr\o m}
\affiliation{Department of Materials, University of California, Santa Barbara, California 93106-9530, USA}
\affiliation{Department of Electrical and Computer Engineering, University of California, Santa Barbara, California 93106-9530, USA}
\author{John M. Martinis}
\affiliation{Department of Physics, University of California, Santa Barbara, California 93106-9530, USA}
\affiliation{California NanoSystems Institute, University of California, Santa Barbara, California 93106-9530, USA}
\author{A. N. Cleland}
\email{anc@physics.ucsb.edu}
\affiliation{Department of Physics, University of California, Santa Barbara, California 93106-9530, USA}
\affiliation{California NanoSystems Institute, University of California, Santa Barbara, California 93106-9530, USA}
\date{\today}

\begin{abstract}
We describe the fabrication and measurement of microwave coplanar waveguide resonators with internal quality factors above $10^7$ at high microwave powers and over $10^6$ at low powers, with the best low power results approaching $2\times10^6$, corresponding to $\sim 1$ photon in the resonator. These quality factors are achieved by controllably producing very smooth and clean interfaces between the resonators' aluminum metallization and the underlying single crystal sapphire substrate. Additionally, we describe a method for analyzing the resonator microwave response, with which we can directly determine the internal quality factor and frequency of a resonator embedded in an imperfect measurement circuit. 
\end{abstract}
\maketitle

High quality factor microwave resonators provide critical elements for superconducting electromagnetic radiation detectors \cite{MKID}, quantum memories \cite{Hofheinz2009, Wang2009}, and quantum computer architectures \cite{Matteo2011}. Good performance and stability can be achieved for such applications using aluminum resonators patterned on sapphire substrates.  Aluminum is a favored material due to its robust oxide and reasonable transition temperature, and sapphire provides an excellent substrate due to its very low microwave loss tangent\cite{Creedon2011} $\delta \sim 10^{-8}$ and its chemical inertness.  However, the quality factors measured in such resonators is lower than expected; recent simulations \cite{Wenner2011} and experiments \cite{Wisbey2010} suggest that the unexplained loss arises mostly from imperfections at the metal-substrate interface. Using an experimental apparatus with minimal stray magnetic fields and infrared light at the sample \cite{Barends2011}, here we show that careful substrate preparation and cleaning yields aluminum-on-sapphire resonators with significantly higher internal quality factors $Q_i$. We also introduce a new method for evaluating the resonator microwave response.

The aluminum for the resonators was deposited on \textit{c}-plane sapphire substrates in one of three deposition systems: A high vacuum DC sputter system (base pressure $P_{\textrm{base}}=6\times10^{-8}$ Torr), a high vacuum electron-beam evaporator ($P_{\textrm{base}}=5\times10^{-8}$ Torr) or an ultra-high vacuum (UHV) molecular beam epitaxy (MBE) system ($P_{\textrm{base}}=6\times10^{-10}$ Torr) with electron-beam deposition. The sapphire substrates were first sonicated in a bath of acetone then isopropanol followed by a deionized water rinse. For the sputter-deposited and e-beam evaporated samples, we further cleaned the substrates prior to Al deposition by Ar ion-milling. For the MBE-deposited samples, we first cleaned the substrates with a load-lock outgassing at 200$^{\circ}$C followed by heating to 850$^\circ$C, in UHV or in \TIL$10^{-6}$ Torr of activated oxygen (O$^{\ast}_{2}$) generated from a radio-frequency plasma source. After substrate treatment, \TIL100 nm of Al was deposited at room temperature. Specific process parameters are given in Table \ref{TabResult}.

\begin{table}[t]
\caption{Sample process information; $w$ is the resonator center stipline width, $f_0$ the resonant frequency, and $Q_i$-H and $Q_i$-L the internal quality factors at high power (before over-saturation) and low power ($\left\langle n_{photon} \right\rangle \TIL1$), respectively.}
\begin{tabular}{l c c c d d}
\hline
\hline
Process\footnote{All films deposited at room temperature.} & \textit{In vacuo}  & $w$  & $f_0$ & \multicolumn{1}{c}{$Q_i$-H} & \multicolumn{1}{c}{$Q_i$-L}\\
& cleaning & ($\mu$m) & (GHz) & \multicolumn{1}{c}{$\times 10^6$} & \multicolumn{1}{c}{$\times 10^6$} \\
\hline
\hline
\multirow{2}{*}{\textbf{A}) Sputter} & 100 eV Ar$^+$ mill  & 3 & 3.833 & 4.3 & 0.16 \\
  &  for 2 min & 15 & 6.129 & 4.5 & 0.30 \\
\hline
\multirow{2}{*}{\textbf{B}) E-beam} & 60 eV Ar$^+$ mill   & 3 & 3.810 & 9.9 & 0.66 \\
  &  for 10 sec & 15 & 6.089 & 4.4 & 0.72 \\
\hline
\multirow{2}{*}{\textbf{C}) MBE} & \multirow{2}{*}{None} & 6 & 4.973 & 5.70 & 0.53 \\
  &  & 15 & 6.120 & 4.33 & 0.76 \\
\hline
\multirow{2}{*}{\textbf{D}) MBE} & \multirow{2}{*}{LL\footnotemark[2] anneal} & 3 & 3.773 & 6.58 & 0.75  \\
  &  & 15 & 6.125 & 5.38 & 0.80 \\
\hline
\multirow{2}{*}{\textbf{E}) MBE} & \multirow{2}{*}{LL\footnotemark[2] \& 850$^{\circ}$C anneal} & 3 & 3.876 & 10.1 & 1.15  \\
  &  & 15 & 6.127 & 6.4 & 0.92 \\
\hline
\multirow{2}{*}{\textbf{F}) MBE} &  LL\footnotemark[2] \& 850$^{\circ}$C anneal & 3 & 3.767 & 12.7 & 1.1 \\
  & in O$^{\ast}_{2}$ \footnotemark[3] & 15 & 6.121 & 8.5 & 1.72 \\
\hline
\hline
\footnotetext[2]{200$^{\circ}$C anneal in load lock (\TIL$10^{-6}$ Torr).}
\footnotetext[3]{Activated oxygen (O$^{\ast}_{2}$) generated from a radio frequency plasma source.}
\end{tabular}
\label{TabResult}
\end{table}

\begin{figure}
\begin{center}
\includegraphics{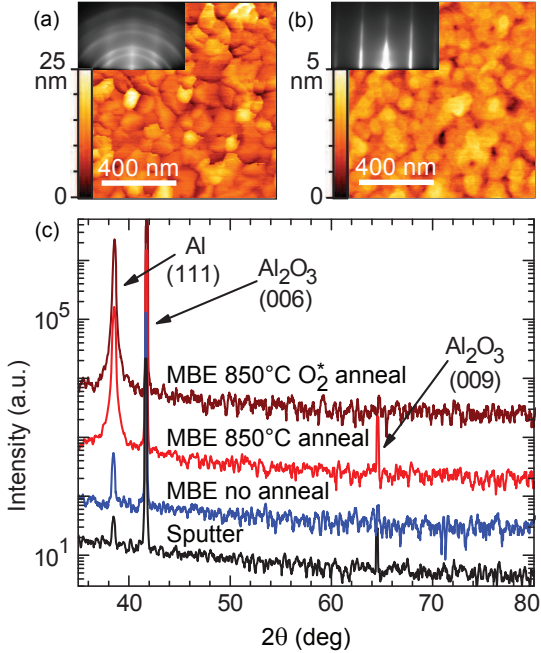}
\end{center}
\caption{(Color online) (a) MBE-deposited Al with no pre-deposition substrate cleaning yields a rough Al surface, shown in the atomic force microscope (AFM) surface scan with $\sim$25 \AA~ RMS roughness and (inset) a diffuse polycrystalline RHEED pattern, compared to (b) MBE-deposited Al with an 850$^\circ$C anneal in O$^{\ast}_{2}$, with a much smoother film ($\sim$4 \AA~ RMS roughness) with sharply streaked RHEED.  (c) $\theta-2\theta$ x-ray diffraction scans of sputtered and MBE-grown films. Vertical scale corresponds to sputtered sample, with other samples plotted on same scale but vertically offset for clarity.  All samples display substrate peaks and an Al(111) diffraction peak.\label{FigMaterialProp}}
\end{figure}

The substrate cleaning process changes the morphology of the deposited Al film, as shown in Fig. \ref{FigMaterialProp}: No cleaning yields a rough Al film and a diffuse polycrystalline reflected high-energy electron diffraction (RHEED) pattern, while a high-temperature anneal gives a much smoother film and sharp crystalline RHEED.  X-ray diffraction scans of all films show strong sapphire substrate peaks as well as an Al(111) peak [Fig. \ref{FigMaterialProp}(c)], with the MBE-grown samples displaying stronger Al peaks. MBE-grown samples heated to 850$^\circ$C with or without O$^{\ast}_{2}$, and MBE samples that underwent only a 200$^\circ$C anneal (not shown), had RHEED, atomic-force microscopy, and x-ray diffraction scans that were nearly indistinguishable from one another.

We patterned resonators from the Al films using an optically-patterned resist and etching in a 300 W BCl$_3$/Cl$_2$ inductively-coupled plasma.  This step defined sets of 12 quarter-wave coplanar waveguide resonators capacitively-coupled to a single central transmission line \cite{Barends2011}.  Resonators were designed with one of three center trace widths: $w$ = 3 $\mu$m, 6 $\mu$m or 15 $\mu$m. The center trace gap to the ground planes on either side was $g=2w/3$. The resonator coupling to the central transmission line corresponded to one of the coupling strengths \cite{Mazinthesis} $Q_c$ = $5\times10^4$, $2\times10^5$, $5\times10^5$ or $1\times10^6$; resonator frequencies $f_0$ ranged from 4 to 8 GHz.

After etching, we stripped the resist and diamond saw-cut the wafers into $6 \times 6$ mm$^2$ dies, placing individual dies in an Al sample box and wiring with 25 $\mu$m dia. Al wire-bonds.  We mounted the sample box on the 50 mK stage of an adiabatic demagnetization refrigerator, enclosed in a copper light-tight shield, surrounded by a magnetic shield at 4 K, as described in Ref. \onlinecite{Barends2011}. We note that all screws and coaxial connectors were made of non-magnetic materials. The use of non-magnetic materials inside of the 4 K magnetic shield reduced the residual magnetic field at the sample from \TIL100 mG to 3 mG, which we verified using cryogenic tests similar to Ref. \onlinecite{Wang2009}.

We measured the transmission coefficient $S_{21}$ of the 50 mK resonators with a vector network analyzer (VNA). The cryostat microwave input cable had 40 dB attenuation at room temperature and 30 dB attenuation at the 4 K stage, followed by 50 mK microwave powder filters \cite{Barends2011} with \TIL1 dB attenuation before the connection to the sample's central transmission line.  The sample's output passed through a second 50 mK microwave powder filter followed by a circulator at 4 K, which protects the resonators from noise originating in the high electron mobility transistor (HEMT) amplifier (noise temperature \TIL4.5 K) also located at 4 K.

\begin{figure*}
\begin{center}
\includegraphics{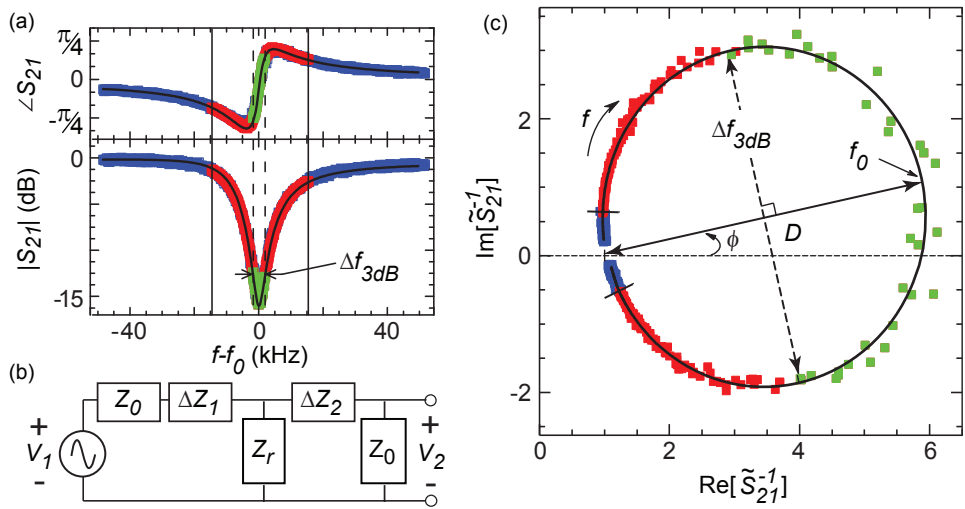}
\end{center}
\caption{(Color online) (a) Measured normalized transmission magnitude $|S_{21}|$ and phase $\angle S_{21}$ (colored squares), with a fit to Eq.\,(\ref{EqS21-1}) (solid black line) which yields $Q_i=1.7\times10^6$, $Q_c^{\ast}=4\times10^5$, $f_0=6.121$ GHz and $\phi=11.7^{\circ}$. The resonator had a $w = 15~\mu$m center stripline width and was patterned from an MBE-deposited Al film on a sapphire substrate that was annealed at 850$^{\circ}$C in O$^{\ast}_{2}$.  The data was taken at 50 mK at low power with $\left\langle n_{photon} \right\rangle \sim 1$ in the resonator.  (b) Circuit diagram including mismatched complex impedances $\Delta Z_1$ and $\Delta Z_2$ for the transmission line input and output, $Z_r$ includes the resonator and its coupling capacitance to the transmission line, which has characteristic impedance $Z_0=50~\Omega$. (c) Parametric plot and fit of Im$[\tilde{S}_{21}^{-1}]$ vs. Re$[\tilde{S}_{21}^{-1}]$ of the same data and fit as (a). Points in the frequency range between the dashed lines (green squares) in (a) correspond to the points to the right of the $\Delta f_{3 \text{dB}}$ dashed line in (c).}
\label{FigMeasurement}
\end{figure*}

A typical normalized transmission spectrum is shown in Fig. \ref{FigMeasurement}(a), displaying a dip in $|S_{21}|$ and a step in the phase of $S_{21}$ near the resonance frequency $f_0$. Note the slight asymmetry about $f_0$ in both the magnitude and phase, which we attribute to a small impedance mismatch in the central transmission line on either side of the resonator\cite{Khalil2011}, likely originating from the wire-bond connections \cite{wirebonds}, sample mount imperfections, or the transmission line geometry. Analyzing a circuit that includes small in-line complex impedances $\Delta Z_1$ and $\Delta Z_2$ on either side of the resonator [Fig. \ref{FigMeasurement}(b)], the transmission $S_{21}$ is given by
\begin{equation}\label{EqS21}
    S_{21}=\frac{2 V_{2}}{V_{1}}=\frac{2Z_0}{Z_1+Z_2} \frac{1}{1+Z/2 Z_{r}},
\end{equation}
where $Z_0 = 50~\Omega$ is the cable impedance, $Z_1 = Z_0 + \Delta Z_1$, $Z_2 = Z_0 + \Delta Z_2$, $1/Z \equiv 1/2Z_1 + 1/2Z_2$, and the impedance $Z_{r}$ of the resonator in series with its coupling capacitor is given by \cite{Mazinthesis}, for frequencies $f$ near resonance,
\begin{equation}\label{EqRatioMS}
    Z_{r}=\frac{Z_0Q_c}{2Q_i}(1+i2Q_i\delta x),
\end{equation}	
with $\delta x = \left(f-f_{0}\right)/f_{0}$, and $Q_{i,c}$ are the internal and coupling quality factors, respectively. The nearly frequency-independent pre-factor in Eq. (\ref{EqS21}), of order unity, is absorbed in an off-resonance transmission calibration of the cabling and amplifiers, leaving us with a normalized inverse transmission $\tilde{S}_{21}^{-1}$ given by
\begin{equation}\label{EqS21-1}
    \tilde{S}_{21}^{-1} = 1 + \frac{Z}{2 Z_{r}} = 1+\frac{Q_i}{Q_{c}^{\ast}}e^{i\phi}\frac{1}{1+i2Q_i\delta x},
\end{equation}
with the magnitude of $Z = |Z| e^{i \phi}$ absorbed in the re-scaled coupling quality factor $Q_{c}^{\ast} \equiv (Z_0/|Z|) Q_{c}$.

In Fig. \ref{FigMeasurement}(c) we display the data in Fig. \ref{FigMeasurement}(a) in the inverse transmission $\tilde{S}_{21}^{-1}$ complex plane. In this representation the resonator response is a circle starting and ending at $\tilde{S}_{21}^{-1}=1$, achieved for frequencies far from resonance. At resonance $f=f_0$, $\tilde{S}_{21}^{-1} = 1+ D e^{i \phi}$ is diametrically opposite $\tilde{S}_{21}^{-1} = 1$,  with diameter $D=Q_i/Q_{c}^{\ast}$ and the circle center rotated from the real axis by the impedance mismatch angle $\phi$. The two frequency points $\pm 90^{\circ}$ around the circle from the resonance frequency are the $\tilde{S}_{21}^{-1}$ 3 dB points, with their frequency difference $\Delta f_{3\text{dB}}$ the full width at half maximum (FWHM) of $|\tilde{S}_{21}^{-1}|$, yielding the internal quality factor $Q_i=f_0/\Delta f_{3\text{dB}}$. Eq. (\ref{EqS21-1}) and the plot of Fig. \ref{FigMeasurement}(c) not only allow the internal quality factor $Q_i$ to be determined directly, these also emphasize how to properly measure and visually scrutinize the data.

\begin{figure}
\begin{center}
\includegraphics{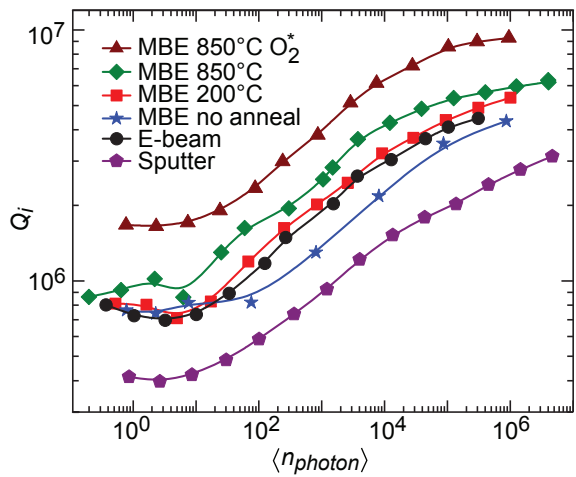}
\end{center}
\caption{(Color online) Power dependence of the internal quality factor $Q_i$ versus average photon number in the resonator $\left\langle n_{photon} \right\rangle$, for resonators with $w = 15~\mu$m. Lines are guides to the eye.  The typical low-power statistical error from a least-squares fit of Eq.\,(\ref{EqS21-1}) is \TIL3\%, smaller than the symbol size.}
\label{FigPower}
\end{figure}

The low-power $Q_i$'s of 15 $\mu$m center trace width resonators [Fig. \ref{FigPower}] range from $3 \times 10^5$ for the sputtered film with ion mill substrate cleaning to $1.7 \times 10^6$ for the MBE-deposited resonator shown in Fig. 2.  We find that as the measurement microwave power is increased [Fig. \ref{FigPower}], the internal quality factors remain roughly proportional, with $Q_i$ for most resonators increasing by about a factor of ten between high and low powers.  The increase in $Q_i$ with power is consistent with loss being dominated by interfacial and surface two-level-states (TLS)\cite{Wenner2011,Sage2011}.

We discovered that for the highest quality factor resonators at the lowest powers, the internal quality factor can fluctuate in time by as much as 30\% over a period of several hours, a variation significantly larger than the statistical uncertainty of the fit $Q_i$. We believe this is due to fluctuations in the population of TLS, which we are actively studying. The values for $Q_i$ reported in this Letter are long-term averages rather than the highest transient values, and are representative of many different samples.

Most significantly, we see a systematic dependence of $Q_i$ on the substrate cleaning and deposition process. We attribute this dependence to changes in the Al-sapphire interface, in agreement with simulations \cite{Wenner2011}. The sputtered film resonators had the lowest $Q_i$ for all powers, and are comparable to the published literature \cite{Wang2009, Sage2011, Khalil2011}. We believe the lower $Q_i$ measured here is due to the pre-deposition ion mill, which may remove most surface contaminants, but may also bury some contaminants as well as leave a damaged substrate surface \cite{Day1993}.  The ion mill used for e-beam evaporation was lower in energy and duration than that for the sputtered films, and produced resonators with $Q_i$'s comparable to those deposited by MBE with no substrate anneal.

We obtained better resonator quality factors with MBE-grown Al with pre-deposition annealing, consistent with better surface properties as shown by RHEED and AFM [Fig. \ref{FigMaterialProp}]. However, the very best quality factors were achieved with high temperature (850$^{\circ}$C) anneals in O$^{\ast}_{2}$, yielding significantly higher $Q_i$ even though RHEED and AFM are nearly identical to other MBE-annealed samples. These results are consistent with microscopic theories attributing TLS to surface-bound hydroxyl groups \cite{Phillips1987}, which can saturate the sapphire (0001) surface \cite{Rabalais1997}, and remain even after annealing at 1100$^{\circ}$C in UHV \cite{Rabalais1997, Niu2000}.  Carbon and hydrogen have also been observed on sapphire after similar annealing processes \cite{Niu2000}.  The highly reactive activated oxygen used here may remove such surface-bound molecules, more effectively than a simple anneal, and with less damage than with ion milling, although we cannot at present verify these hypotheses.

In summary, we have measured the internal quality factors of superconducting aluminum resonators while varying the substrate preparation, deposition method and measurement power.  Cleaning the substrate surface as thoroughly as possible without damaging the underlying crystal leads to the highest $Q_i\sim 1.7 \times 10^6$ at single photon energies, achieved for resonators with $w=15~\mu$m and $f_0 \cong 6$ GHz. The same process also produces the best high-power $Q_i \sim 12.7 \times 10^7$, but for resonators with $w=3~\mu$m and $f_0 \cong 4$ GHz.

The authors thank B. A. Mazin for stimulating discussions. Devices were fabricated at the UCSB Nanofabrication Facility, a part of the NSF-funded National Nanotechnology Infrastructure Network.  M. M. acknowledges support from an Elings Postdoctoral Fellowship.  This work was supported by IARPA under ARO award W911NF-09-1-0375.


\begin{thebibliography}{18}

\bibitem{MKID} P. K. Day, H. G. LeDuc, B. A. Mazin, A. Vayonakis, and J. Zmuidzinas, Nature (London) \textbf{425}, 817 (2003).

\bibitem{Hofheinz2009} M. Hofheinz, H. Wang, M. Ansmann, R. C. Bialczak, E. Lucero, M. Neeley, A. D. O'Connell, D. Sank, J. Wenner, J. M. Martinis, A. N. Cleland, Nature (London) \textbf{459}, 546-549 (2009).

\bibitem{Wang2009} H. Wang, M. Hofheinz, J. Wenner, M. Ansmann, R. C. Bialczak, M. Lenander, E. Lucero, M. Neeley, A. D. O'Connell, D. Sank, M. Weides, A. N. Cleland, and J. M. Martinis, Appl. Phys. Lett. \textbf{95}, 233508 (2009).

\bibitem{Matteo2011} M. Mariantoni, H. Wang, T. Yamamoto, M. Neeley, R. C. Bialczak, Y. Chen, M. Lenander, E. Lucero, A. D. O'Connell, D. Sank, M. Weides, J. Wenner, Y. Yin, J. Zhao, A. N. Korotkov, A. N. Cleland, J. M. Martinis, Science \textbf{334}, 61 (2011).

\bibitem{Creedon2011} D. L. Creedon, Y. Reshitnyk, W. Farr, J. M. Martinis, T. L. Duty, M. E. Tobar, Appl. Phys. Lett. \textbf{98}, 222903 (2011).

\bibitem{Wenner2011} J. Wenner, R. Barends, R. C. Bialczak, Y. Chen, J. Kelly, E. Lucero, M. Mariantoni, A. Megrant, P. J. J. O'Malley, D. Sank, A. Vainsencher, H. Wang, T. C. White, Y. Yin, J. Zhao, A. N. Cleland, and J. M. Martinis, Appl. Phys. Lett. \textbf{99}, 113513 (2011).

\bibitem{Wisbey2010} D. S. Wisbey, J. Gao, M. R. Vissers, F. C. S. da Silva, J. S. Kline, L. Vale, and D. P. Pappas, J. Appl. Phys. \textbf{108}, 093918 (2010).

\bibitem{Barends2011} R. Barends, J. Wenner, M. Lenander, Y. Chen, R. C. Bialczak, J. Kelly, E. Lucero, P. J. J. O'Malley, M. Mariantoni, D. Sank, H. Wang, T. C. White, Y. Yin, J. Zhao, A. N. Cleland, J. M. Martinis, and J. J. A. Baselmans, Appl. Phys. Lett. \textbf{99}, 113507 (2011).

\bibitem{Mazinthesis} B. A. Mazin, Ph.D. thesis, California Institute of Technology, 2004.

\bibitem{Khalil2011} M. S. Khalil, F. C. Wellstood, K. D. Osborn, IEEE Transactions on Applied Superconductivity \textbf{21}, 879 (2011).

\bibitem{wirebonds} G. J. Grabovskij, L. J. Swenson, O. Buisson, C. Hoffmann, A. Monfardini, and J.-C. Vill\'{e}gier, Appl. Phys. Lett. \textbf{93}, 134102 (2008).

\bibitem{Sage2011} J. M. Sage, V. Bolkhovsky, W. D. Oliver, B. Turek, and P. B. Welander, J. Appl. Phys. \textbf{109}, 063915 (2011).

\bibitem{Day1993} M. E. Day, M. Delfino, W. Tsai, A. Bivas, and K. N. Ritz, J. Appl. Phys. \textbf{74}, 5217 (1993).

\bibitem{Phillips1987} W. A. Phillips, Rep. Prog. Phys. \textbf{50}, 1657 (1987).

\bibitem{Rabalais1997} J. Ahn, J. W. Rabalais, Surf. Sci. \textbf{388}, 121 (1997).

\bibitem{Niu2000} C. Niu, K. Shepherd, D. Martini, J. Tong, J. A. Kelber, D. R. Jennison, and A. Bogicevic, Surf. Sci. \textbf{465}, 163 (2000).






\end{thebibliography}
\end{document}